\documentstyle[prl,aps,preprint]{revtex}
\begin{document}
\draft

\title{Resonant Tunneling and Coulomb Oscillations}

\author{J\"urgen K\"onig$^1$, Herbert Schoeller$^{1,2}$, and Gerd Sch\"on$^1$}

\address{
$^1$ Institut f\"ur Theoretische Festk\"orperphysik, Universit\"at
Karlsruhe, 76128 Karlsruhe, Germany\\
$^2$ Department of Physics, Simon Fraser University, Burnaby, B.C.,
V5A 1S6, Canada}

\date{\today}

\maketitle

\begin{abstract}
The influence of quantum fluctuations on electron transport through small
metallic islands with Coulomb blockade effects is studied beyond the
perturbative regime.
In tunnel junctions with low resistance higher order coherent
processes and ``inelastic resonant tunneling'' become important.
We present a path integral real-time description, which allows a systematic
diagrammatic classification of these processes.
Quantum fluctuations renormalize system parameters and
lead to finite lifetime broadening. Both effects are observable
in the gate voltage dependence of the nonlinear conductance. The
finite bias voltage introduces an energy scale up to which
quantum fluctuations are probed. It can be larger
than the temperature, which makes quantum effects experimentally observable
above thermal fluctuations.
\end{abstract}

Electron transport through mesoscopic metallic islands
is strongly influenced by the charging energy
associated with the low capacitance of the junctions
\cite{Ave-Lik,Gra-Dev,Schoen-Zai}.
A variety of single-electron phenomena, including Coulomb blockade
at low temperatures and Coulomb oscillations of the conductance
as a function of a gate voltage have been observed.
A master equation description of sequential tunneling
\cite{Ave-Lik} is sufficient as long as
the conductance of the barriers is low,
$\alpha_0\equiv h/(4\pi^2 e^2 R_T) \ll 1$.
In general, quantum fluctuations and higher order
coherent tunneling processes should be considered
\cite{Laf-exp,Gla-Mat,Zai,Gra,Fal-Scho-Zim,SS}.
This includes ``cotunneling''\cite{Av-Naz}, where in a second order coherent
process electrons tunnel via a virtual state of the island.
Furthermore, resonant tunneling plays a role.
In comparison to the well known case of independent electrons
we encounter here two complications:
(i) The metallic system contains many electrons, and
with overwhelming probability different electron states are involved in the
different transitions of the coherent process (this is denoted by
``inelastic'').
(ii) The strong Coulomb interaction cannot be accounted for perturbatively.

In the present article we develop a systematic diagrammatic technique to
identify the processes of sequential tunneling, cotunneling and inelastic
resonant tunneling. We study the time evolution of the density matrix.
In an earlier paper \cite{SS} we had formulated the problem, after a
separation of charge and fermionic degrees of freedom, in a many-body
expansion technique.
Here we reformulate it in a real-time path-integral representation
\cite{kssf}.
The latter is familiar from the work of Caldeira and Leggett \cite{CLeggett}
and Feynman and Vernon \cite{Fey-Ver}
who studied dissipation in quantum mechanics.
Dissipation associated with tunneling of electrons was investigated in
Refs. \cite{Eck-Scho-Amb,Schoen-Zai}.
Similar problems arise in the context of local, strongly correlated Fermi
systems like the Kondo and Anderson model\cite{Bic}.
An essential step in the present work is a transformation of the
path-integral description of electron tunneling from a phase
to a charge representation \cite{Schoen-Zai,kssf}.

As examples we study the single electron transistor
\cite{Gra-Dev} and the electron box.
In the transistor a metallic island is
coupled by two tunnel junctions ($L,R$) to two electrodes,
and further capacitively to a gate voltage $V_G$.
A transport voltage $V = V_L - V_R$ drives a current.
The charging energy is $E_{ch}(n) = (ne-Q_G)^2/(2C)$. It
depends on the number of excess electrons $n$ on the island and
on the continuously varying external charge
$Q_G = C_G V_G + C_L V_L + C_R V_R$.
The total island capacitance
$C = C_L + C_R + C_G$
defines the scale of the charging energy, $E_C \equiv e^2/2C$.
The electron box,
consisting of an island coupled via a tunnel junction
to one electrode and capacitively to the gate voltage, is described
similarly.

At low temperature the electron number $n(Q_G)$ in the electron box
increases in unit steps by tunneling as $Q_G$ is increased.
In the transistor tunneling occurs in lowest order
perturbation theory only if the electrochemical potential of one
electrode is high enough to allow an electron to enter the island,
say $eV_L > E_{ch}(n+1) - E_{ch}(n) $, while the electrochemical
potential of the other electrode allows the
tunneling process to that electrode, i.e.
$E_{ch}(n+1) - E_{ch}(n) > eV_R$.
Within the window set by these conditions the current  is
$4R_t I(V,Q_G) = V - 4 [Q_G-(n+1/2)e]^2/(C^2V)$.
I.e. the nonlinear conductance, $G(V,Q_G) = \partial{I}/\partial{V}$,
shows as a function of $Q_G$ an $e$-periodic series of structures
of width $CV$ with, at $T=0$, vertical steps at its edges.
At finite temperature $\langle n(Q_G) \rangle $ and
the steps are washed out.

Quantum fluctuations further wash out these steps.
Below we describe these processes diagrammatically
and determine their effect on the nonlinear conductance.
In the most interesting case we can re-sum the diagrams
and obtain closed expressions for the stationary density matrix and
the spectral density describing the charge excitations of the system.
Our main findings are:
(i) Quantum fluctuations renormalize the
energy and conductance.
(ii) The coherent resonant tunneling processes
 give rise to a broadening of the charge state levels.
The effects are observable experimentally in the nonlinear conductance.
Here the finite bias voltage introduces an energy scale,
up to which quantum fluctuations are probed.
Since $eV$ can easily be chosen larger than the temperature,
quantum effects are not hidden below thermal broadening and
become observable in a realistic experiment.

The description of the single electron transistor
is based on the Hamiltonian $	H=H_L+H_R+H_I+H_{ch}+H_{t,L} +H_{t,L} $.
The first terms describe noninteracting electrons in the
metallic left and right lead and island.
These are treated as reservoirs,
i.e. their electron distributions are assumed to be thermal
with electrochemical potential which depends on the state. The Coulomb
interaction $H_{ch}$ depends on the charge on
the island, as expressed by the charging energy  $E_{ch}(n)$ given above.
Charge transfer is described by standard tunneling Hamiltonians
$	H_{t,r}=\sum_{k\in r,q\in I,\sigma} T_{r}
	c^{\dagger}_{k\sigma} c_{q\sigma} + h.c.\; , $
where $r = L,R$. The matrix elements and densities of states
determine the tunnel conductances,
$R^{-1}_{r} = (4\pi e^2 /\hbar) N_{r}(0) N_I(0) |T_{r}|^2 $.
We consider ``wide'' metallic junctions with many
transverse channels $N_{ch} \gg 1$.
Hence ``inelastic'' higher order tunneling processes, involving
different electron states for each step, dominate over ``elastic''
processes which involve the same state repeatedly.

We describe the system of interacting electrons
in a path integral representation.
A Hubbard-Stratonovich transformation, used to handle the capacitive
interaction, introduces as collective variable
the phase $\varphi$, the quantum
mechanical conjugates
of the charge on the island.
The phases in the electrodes are
fixed by the applied voltages, $\varphi_r=eV_r t$.
Next, the electronic degrees of freedom can be traced out,
followed by an expansion of the electron propagators.
Since we consider wide junctions, only simple loops need to be retained
\cite{bruder}. Their iteration introduces in each
order a factor  $N_{ch}$,
hence they dominate over more complicated higher order loops.
After this stage the system is described by a reduced density matrix
$\rho(\{\varphi_{1}\},\{\varphi_{2}\})$. It can be expressed
by an effective action, which depends on the phases $\varphi_\sigma$
corresponding to the forward and backward propagator $\sigma =1,2$.
The structure of the theory is familiar from Refs.
 \cite{Fey-Ver,CLeggett}, where a quantum system coupled to a harmonic
 oscillator bath has been considered, and from
 Refs. \cite{Eck-Scho-Amb,Schoen-Zai}, where electron tunneling has been
 described.
The effective action contains the charging energy of the system
appropriate to $E_{ch}$.
The tunneling couples the forward and backward
time propagators. For each junction we have
 \cite{Eck-Scho-Amb,Schoen-Zai}
\begin{equation}
	S_{t,r}[\varphi_{1},\varphi_{2}] =
	4 \pi i \sum_{\sigma,\sigma'=1,2}
	\int_{t_i}^{t_f} dt
	\int_{t_i}^{t} dt' \alpha^{\sigma,\sigma'}_r(t-t')
	\cos[\varphi_{\sigma}(t) - \varphi_{\sigma'}(t')] \; .
\end{equation}
The kernels are given in Fourier space by
$\alpha^{\sigma,1}_r(\omega) = (-1)^{\sigma+1} \alpha^-_r (\omega)$,
$\alpha^{\sigma,2}_r(\omega) = (-1)^{\sigma} \alpha^+_r (\omega) $
with $ \alpha^{\pm}_r(\omega)=
    \pm\alpha_{0,r}\,(\omega-eV_r) \,
[\exp(\pm\beta (\omega-eV_r)-1]^{-1} $.

An important step for a systematic description of tunneling
processes is the change from the phase to a charge
representation\cite{Schoen-Zai}, accomplished by ($\hbar=k=1$)
\begin{eqnarray}
        &&\rho(t_f;n_{1f},n_{2f}) = \nonumber\\
	&&\sum_{n_{1i},n_{2i}}\rho(t_i;n_{1i},n_{2i})
        \int d\varphi_{1f} d\varphi_{2f} d\varphi_{1i} d\varphi_{2i}
	\int_{\varphi_{1i}}^{\varphi_{1f}} \cal{D}\varphi_1(t)
        \int_{\varphi_{2i}}^{\varphi_{2f}} \cal{D}\varphi_2(t)
	\int \cal{D} n_1(t) \int \cal{D} n_2(t) \nonumber\\
	&&\exp \left\{i\sum_r S_{t,r}[\varphi_1,\varphi_2]
	+ \sum_{\sigma=1,2}(-1)^\sigma
	 \left(in_{\sigma i}\varphi_{\sigma i}-in_{\sigma f}\varphi_{\sigma f}
	+iS_{ch}[n_\sigma]-i\int d t \, n_\sigma \dot{\varphi}_\sigma \right)
	 \right\} \; .
\end{eqnarray}
The charging energy is
$S_{ch}[n_{\sigma}] = \int_{t_i}^{t_f} dt \frac{1}{2C}(n_{\sigma}e-Q_G)^2 $.
In systems with discrete charges the integrations
include a summation over winding numbers \cite{Schoen-Zai}.
Next we expand the tunneling terms $\exp(iS_t)$ and
integrate over $\varphi_{\sigma}$.
Each of the exponentials $\exp[\pm i\varphi_{\sigma}(t)]$
describes tunneling of an electron
at time $t$ on the forward or backward branch, $\sigma = 1$ or $2$,
respectively.
These changes occur in pairs in each junction $r=L,R$ and are connected by
$\alpha^{\sigma,\sigma'}_r(t-t')$.
Each term of the expansion can be visualized by a diagram.
Several examples are displayed in Fig. (\ref{fig2}).
There is a closed time-path consisting of two horizontal lines,
corresponding to the forward and backward propagator
between $t_i$ to $t_f$.
Along the time-path vertices are arranged,
connected in pairs by (dashed) tunneling lines, either within one
propagator or between the two propagators.

We start from a density matrix which is diagonal,
 $\rho(t_i) = P^{(0)}(n) |n\rangle\langle n|$. During transitions
 the system is in an off-diagonal state.
We denote the sum of all diagrams, starting and ending in the diagonal
states $n$ and $n'$, respectively, by $\Pi_{n,n'}$.
It can be expressed by an irreducible self-energy
$\Sigma_{n,n'}$, which describes the transitions,
in the style of a Dyson equation
\begin{equation}
	\Pi_{n,n'} =
	{\Pi^{(0)}}_{n} + \sum_{n''}
	\Pi_{n,n''}\, \Sigma_{n'',n'} \,
	{\Pi^{(0)}}_{n'} \; .
\label{pi}
\end{equation}
Here $\Pi^{(0)}$ is the free propagator.
We identify the solution of Eq. (\ref{pi}), multiplied with
$P^{(0)}(n)$, as the stationary distribution function
$\sum_n P^{(0)}(n)\Pi_{n,n'} = P^{st}(n')$.
The sum rule $\sum_{n^\prime}\Sigma_{n,n^\prime}=0 \,$ (see Ref.
 \cite{kssf} for further details) implies
\begin{equation}\label{31}
	0 = - P^{st}(n) \sum_{n^\prime \ne n}\Sigma_{n,n^\prime}
	+ \sum_{n^\prime \ne n}P^{st}(n^\prime)\Sigma_{n^\prime,n} \;.
\end{equation}
I.e., we recover a stationary master equation.
In general the irreducible self-energy $\Sigma_{n',n}$ yields
the transition rates of all possible correlated tunneling processes.
We, furthermore, note that the stationary solution $P^{st}(n)$
is independent of the initial distribution $P^{(0)}(n)$.

For illustration we evaluate all diagrams which contain no overlapping
tunneling lines, as visualized on the left hand side of Fig. (\ref{fig2}).
In this case the irreducible self-energy parts
\begin{equation}
	\Sigma^{(1)}_{n,n\pm1} = 2\pi i\alpha^\pm (\pm \Delta E_{ch}^\pm)
	\qquad , \qquad
	\Sigma^{(1)}_{n,n} = -2\pi i \sum_\pm\alpha^\pm
	(\pm \Delta E_{ch}^\pm) \; ,
\label{sigma}
\end{equation}
where $\Delta E_{ch}^\pm =E_{ch}(n\pm1)-E_{ch}(n)$, reproduce
the well-known single electron tunneling rates.
In situations where single electron tunneling is suppressed by Coulomb
blockade the lowest order contribution to the current arises due to
cotunneling.
It is described by a diagram, also  shown in Fig. (\ref{fig2}),
with tunneling processes in the left and in the
right junction, where the corresponding lines $\alpha_L(t_L-t_L')$ and
$\alpha_R(t_R-t_R')$ overlap in time.
Applying our rules \cite{kssf} we reproduce the well-known cotunneling rate.

The perturbative approach is sufficient for
$\alpha_0 \ln{({E_C\over 2\pi T})}\ll 1$. At larger values of the
conductance resonant tunneling processes get important.
To proceed we have to find a systematic criterion which diagrams to
retain. For this we
note that during a tunneling process the reservoirs contain an electron
excitation.
Our criterion is to take into account only matrix elements of the
 density matrix which differ at most
by two excitations in the leads or (equivalently) in the island.
This means that in the diagrams any vertical line cuts at most
two tunneling lines.
We, furthermore, concentrate on situations
where only two charge states, $n=0, 1$, need to be considered.
This is sufficient when the
energy difference of the two states
$\Delta_0 \equiv E_{ch}(1)-E_{ch}(0)$ and
the bias voltage $eV=eV_L-eV_R$
are low compared to the energy, $E_C$,
associated with transitions to higher states.
The combination of both restrictions
implies that the diagrams contain no crossing tunneling lines,
which allows us to evaluate the irreducible self-energy analytically.

Using the notations
$\alpha_r(\omega)=\alpha^+_r(\omega)+\alpha^-_r(\omega)$,
$\alpha(\omega)=\sum_r\alpha_r(\omega)$
and $\alpha_0=\sum_r\alpha_{0,r}$ we find
\begin{eqnarray}
	\Sigma_{0,1}=-\Sigma_{0,0}=2\pi i{\lambda_+\over\lambda}
	\qquad &,& \qquad
	\Sigma_{1,0}=-\Sigma_{1,1}=2\pi i{\lambda_-\over\lambda}
\label{40}\\
	\mbox{with}\qquad
\label{39}
	\lambda_\pm=\int d\omega\,\alpha^\pm(\omega)|\pi(\omega)|^2
	\qquad&,& \qquad
	\lambda= \int d\omega \,|\pi(\omega)|^2 ,\\
\label{35}
	\mbox{and}\qquad
	\pi(\omega)=
	{1\over\omega-\Delta_0-\sigma(\omega)}\qquad&,& \qquad
	\sigma(\omega)= -\int d\omega^\prime\,{\alpha(\omega^\prime)
	\over \omega^\prime-\omega-i\eta}.
\end{eqnarray}
Inserting these quantities in Eq. (\ref{31})
we arrive at the stationary, normalized probabilities
$P_0^{st}=\lambda_-$ and $P_1^{st}=\lambda_+\,$,
with $\lambda_+ +\lambda_- =1$.

The expression for the current at time $t$ in the junction $r$ can
be written as
\begin{equation}
	I_r (t) = 4\pi ie  \int^{t}_{-\infty} dt' \sum_{\sigma}
	\alpha^{1,\sigma}_r(t-t')
	\langle \sin[\varphi_{1}(t)-\varphi_{\sigma}(t')]
	\rangle \, ,
\label{current}
\end{equation}
where the expectation value is taken with the density matrix
discussed above, with $t=t_f$. We, therefore, study the correlation
functions describing charge transfer at times $t$ and $t^\prime$
\begin{equation}
	C^>(t,t^\prime)=-i{\langle e^{-i\varphi(t)}e^{i\varphi(t^\prime)}
	\rangle} \qquad , \qquad
	C^<(t,t^\prime)=i{\langle e^{i\varphi(t^\prime)}e^{-i\varphi(t)}
	\rangle} \; .
\label{13}
\end{equation}
We further introduce a spectral density for charge excitations
$	A(\omega)={1\over 2\pi i}[C^<(\omega)-C^>(\omega)] $.
Within the approximations described above we find
\begin{eqnarray}
        C^{^<_>}(\omega)&=&\pm2\pi i \sum_r\alpha_r(\omega)
	f[\pm(\omega-eV_r)]
	|\pi(\omega)|^2 \, , \label{46d}\\
	A(\omega)&=&\alpha(\omega)\,|\pi(\omega)|^2\label{45} \\
\mbox{and} \qquad
	I_r&=&{e\over h}4\pi^2\int d\omega\sum_{r^\prime}
	{\alpha_{r^\prime}(\omega)\alpha_r(\omega)\over\alpha(\omega)}
	A(\omega) [f(\omega - eV_{r^\prime})-f(\omega-eV_r)] \; .
\label{46c}
\end{eqnarray}
These results satisfy conservation laws and sum rules, $\sum_r I_r=0$ and
$\int d\omega A(\omega) =1$, and the equilibrium relations
between correlation functions and the spectral density.
The classical result is recovered
in lowest order in $\alpha_0$, where
$A^{(0)}(\omega)=\delta(\omega-\Delta_0)$.
Quantum fluctuations yield energy renormalization and broadening effects,
which enter in the spectral density via the complex
self-energy $\sigma(\omega)$ given in Eq. (\ref{35}).

In equilibrium, for $V_R=V_L$, the transistor is
equivalent to the single electron box. The average electron
number becomes
$\langle n \rangle=\int d\omega f(\omega) A(\omega)$.
In the classical limit it reduces to
$\langle n^{cl} \rangle =f(\Delta_0)$,
where the energy difference $\Delta_0=E_C(1-2C_GV_G/e)$ depends on
the gate voltage. It shows a
step at $Q_G = e/2$, which is smeared by temperature.
At larger values of $\alpha_0$ or lower temperature we have to include
the self-energy $\sigma(\omega)$ in the spectral density.
The limits $T,V=0$ and $|\omega|\le T,V$ can be
analyzed analytically. In the former,
\begin{eqnarray}
	A(\omega)&\cong&{|\omega|\over\Delta_0}\,\cdot\,
	\frac{ \tilde{\Delta}(\omega) \tilde{\alpha}(\omega)}
	{[\omega-\tilde{\Delta}(\omega)]^2
	+[\pi \tilde{\Delta}(\omega) \tilde{\alpha}(\omega)]^2}\; ,\\
	\tilde{\Delta}(\omega)&=&\frac{\Delta_0}{1+2\alpha_0
	\ln({E_C\over|\omega|})} \cdot
	\frac{1}{1+\pi^2\tilde{\alpha}(\omega)^2}\; , \\
	\tilde{\alpha}(\omega)
	&=&{\alpha_0\over 1+2\alpha_0 \ln({E_C\over
	|\omega|})} \; .
\end{eqnarray}
The spectral density $A(\omega)$ has a maximum at the renormalized
energy difference $\Delta$, obtained from
the self-consistent solution of $\Delta=\tilde{\Delta}(\Delta)$
\cite{Fal-Scho-Zim,Gla-Mat}.
It further is broadened by $\pi\Delta\alpha$,
 where $\alpha=\tilde{\alpha}(\Delta)$.
Due to quantum fluctuations the step of the average charge
$\langle n \rangle$ in the electron box at the degeneracy point $\Delta_0=0$
is washed out even for $T=0$.
At finite $T$ the slope at $\Delta_0=0$ shows an
anomalous temperature dependence,
$\partial \langle n \rangle /\partial\Delta_0|_{\Delta_0=0}
\cong -\{4T[1+2\alpha_0 \ln{({E_C\over 2\pi T})}]^2\}^{-1}$.

A pronounced and experimentally accessible
signature of quantum fluctuations is contained in the
nonlinear response of a transistor. We study the
differential conductance $G(V,Q_G)=\frac{\partial{I(V)}}{\partial{V}}$
at finite voltages.
In this case the spectral density in the limit $|\omega|\le T,V$ is involved.
Here we find a renormalization of $\Delta$ and $\alpha_0$ by a factor $Z$ with
$Z^{-1}=1+2\alpha_0 \ln({E_C / \max \{eV/2,2\pi T\} })$ and a broadening of
the spectral density which is given by $Z \alpha(\omega)$.
Now $eV$ provides an energy scale, and renormalization and life-time effects
are probed over a wide energy range even at zero temperature.
The result of Eq. (\ref{46c}) is plotted in Fig.(\ref{fig7}).
For comparison we also show the classical result,
where the conductance is nonzero only in the range
$|\Delta_0|\le\frac{eV}{2}$, with vertical steps at the edges.
Fig.(\ref{fig7}) clearly displays the renormalization effects
and, moreover, the finite life-time broadening.
The width of the structure as a function of $\Delta_0/V$ is $Z^{-1}$ which
depends logarithmically on $V$.
These effects are observable in an experiment with realistic parameters.
In particular, a transport voltage can be applied which exceeds
the temperature, $kT \ll eV \le E_C$. Hence thermal smearing
is still weak, and quantum effects are visible.

In conclusion, we have presented a systematic description
of coherent single electron tunneling processes
including resonant tunneling in a transistor. We propose to
measure the nonlinear conductance in order to study
renormalization effects and lifetime broadening induced by
these quantum effects.

We acknowledge the discussions
with G. Falci, A.D. Zaikin and G. Zimanyi.
This work was supported by the 'Sonderforschungsbereich 195'
of the DFG and the Swiss National Science Foundation (H.S.).

\begin{figure}
\caption{
Example of a diagram showing various tunneling processes:
on the left sequential tunneling in the left and right junctions,
then a term which preserves the norm, next a cotunneling process,
and on the right resonant tunneling.
}
\label{fig2}
\end{figure}

\begin{figure}
\caption{
The differential conductance at
$T=0$ in the nonlinear response regime as function
of the gap energy normalized to the transport voltage $V$.
We consider a symmetric bias and choose
$\alpha_{0,L}= \alpha_{0,R}=0.05$,
and $(1)\,eV/E_C=0.1\,,\,(2)\,eV/E_C=0.01\,,
\,(3)\,eV/E_C=0.001\,$. For comparison, $(0)$ shows the result obtained
in lowest order perturbation theory.
}
\label{fig7}
\end{figure}

\end{document}